\documentclass[11pt]{article}

\usepackage[margin=1in]{geometry}
\usepackage{setspace}
\usepackage{amsmath,amssymb,amsfonts,mathtools,bm}
\usepackage{booktabs}
\usepackage{tabularx}
\usepackage{array}
\usepackage{natbib}

\newtheorem{theorem}{Theorem}[section]
\newtheorem{corollary}[theorem]{Corollary}
\newtheorem{proposition}[theorem]{Proposition}

\title{Flexible Estimation of Covariate-Specific Win Ratio for
Prioritized Time-to-Event Outcomes}

\author{
Zhiguo Li
\quad
Xiaofei Wang\\[0.5em]
\small Department of Biostatistics and Bioinformatics, Duke University,\\
\small Durham, North Carolina, U.S.A.
}

\date{}

\begin{document}

\maketitle

%\begin{center}
%\small
%\textbf{Running headline:} Covariate-Specific Win Ratio
%\end{center}

\begin{abstract}
Prioritized time-to-event outcomes arise frequently in clinical studies, where treatment comparisons may be based on a prespecified hierarchy of clinically important outcomes. The win ratio provides a natural summary of treatment benefit for such prioritized outcomes. However, methods for characterizing heterogeneous treatment effects by estimating how the win ratio varies with baseline covariates remain limited in this setting. We develop a flexible local inverse probability of censoring weighted estimator of the win ratio as a function of baseline covariates, without imposing the parametric shape or proportional win-fractions structure assumed by existing regression approaches. We use best linear projections to summarize the estimated log win ratio and develop formal tests of log-linearity of the win ratio function. We also propose a complementary model-based estimator of the covariate-specific win ratio assuming proportional hazards models for the marginal distributions, copula dependence between the margins, and Bernstein-polynomial sieves for the baseline hazards. Large-sample properties and bootstrap validity are established for both approaches. Simulations show that the local estimator is robust, while the model-based procedure can provide substantial efficiency and power gains under correct specification, and proportional win-fractions regression may lead to bias when its structural assumptions fail. The methods are illustrated by analyzing data from CALGB 9633, a randomized lung cancer trial.
\end{abstract}


\raggedbottom

\section{Introduction}

Clinical studies increasingly rely on time-to-event outcomes that are naturally prioritized according to clinical importance. For example, in oncology, death is often prioritized over disease progression, and in cardiovascular studies, mortality is prioritized over nonfatal events. Standard composite endpoints such as progression free survival collapse such outcomes into a single event time and therefore obscure the clinical hierarchy among component outcomes. The win ratio is a particularly interpretable generalized pairwise comparison method for prioritized outcomes, comparing patients across treatment groups according to a prespecified clinical hierarchy and summarizing treatment benefit through win and loss probabilities \citep{Buyse2010,Pocock2012}. Subsequent work has clarified the estimand, censoring adjustment, and extensions of win ratio type methods for prioritized time-to-event outcomes \citep[and references cited therein]{Oakes2016,Peron2018,Dong2020,Parner2025}.

Most existing win ratio methods either target marginal treatment comparisons or rely on structured regression models. Marginal and inverse probability of treatment weighted (IPTW) win ratio methods summarize overall or weighted population-level effects, but do not estimate how treatment benefit varies continuously with patient characteristics \citep[e.g.][]{Dong2020,WangIPTW2025}. Stratified analyses \citep{Dong2018,Dong2023} provide subgroup-specific summaries, including for categorized continuous covariates, but require predefined cut points and are limited for studying smooth continuous or multivariable effect modification. The proportional win-fractions regression model of \citet{MaoWang2021} offers an important regression approach, but imposes a global log-linear model and a time-invariant win-fraction structure. This may be restrictive because the win ratio is inherently horizon-specific: \citet{Oakes2016} showed that horizon invariance requires a strong bivariate Lehmann-type structure on the joint survivor functions. Thus, nonlinear, localized, or horizon-dependent covariate-specific treatment heterogeneity may be obscured by existing approaches.

We are partly motivated by the analysis of prioritized outcomes in CALGB 9633, a lung cancer trial considered in Section~\ref{sec:application}, in which time to death is treated as the primary outcome and time to recurrence or progression as the secondary outcome. Previous analyses of CALGB 9633 using disease-free survival suggested possible treatment-effect heterogeneity by tumor size \citep{Strauss2008}, motivating further investigation of heterogeneity under a prioritized outcome framework. At the 5-year horizon, when treatment-effect heterogeneity was examined with respect to age, the proportional win-fractions diagnostic test provided evidence against the proportionality assumption underlying the existing regression approach of \citet{MaoWang2021}. Moreover, the age-specific win ratio exhibited substantial nonlinearity, so that a single global log-linear effect did not adequately describe the covariate-specific treatment comparison. This provides a concrete setting in which a flexible covariate-specific analysis is scientifically useful but the existing proportional win-fractions regression model is difficult to justify. More generally, such settings motivate an approach that estimates the win ratio flexibly as a function of covariates and treats low-dimensional regression summaries as descriptive approximations rather than structural assumptions.

Motivated by this, we develop a local inverse probability of censoring weighted (IPCW) estimator
of the covariate-specific win ratio for prioritized
right-censored time-to-event outcomes, including terminal-event settings such
as death prioritized over progression. Instead of imposing a parametric,
log-linear, or proportional win-fractions assumption, the proposed method
estimates the win ratio as a function of the covariate locally in covariate
space using kernel smoothing ideas for censored survival data
\citep{Beran1981,Dabrowska1987,Mielniczuk1987,WandJones1995,FanGijbels1996}.
This curve-first approach can reveal nonlinear or localized treatment-effect
heterogeneity. To obtain interpretable low-dimensional summaries, we consider
best linear projections of the estimated log win ratio function and formally
assess whether a log-linear representation is adequate. When log-linearity is
not supported, the best linear projection remains meaningful as a summary of
the overall linear trend, while the flexible curve is needed to describe the
nonlinear features of treatment-effect heterogeneity.

Under conditional independent censoring, the win and loss probabilities admit
integral representations in terms of marginal and joint survival functions.
We therefore estimate these survival components locally within treatment
arms, using kernel weights and a local product-limit estimator for censoring,
and plug them into the integral formulas
\citep{Beran1981,Dabrowska1987,LinYing1993}. Unlike the usual pairwise
comparison approaches that classify observed subject pairs and then
adjust incomplete comparisons \citep[e.g.,][]{Pocock2012,Dong2020}, this
formulation estimates the covariate-specific win and loss probabilities
directly and separates the target parameter from the handling of incomplete pairwise
orderings. It also suggests a path toward interval-censored outcomes by
replacing the right-censored survival estimators with suitable
interval-censored counterparts.

Furthermore, the integral representations of the win and loss probabilities
suggest a natural model-based approach to covariate-specific win ratio
estimation through models for the conditional survival distributions of the
component event times. Proportional hazards (PH) models provide a common
choice for the marginal survival distributions, while their dependence can
be represented through a copula \citep{Nelsen2006}. To retain flexibility in
the marginal event-time distributions, the unknown baseline cumulative
hazards can be estimated semiparametrically using Bernstein-polynomial sieves
\citep{ZhouHuSun2017,SunDing2021}. This leads to a complementary model-based
estimator that can be substantially more efficient than the nonparametric
local estimator when the working models are correctly specified. Importantly, the assumptions underlying the PH model-based approach are different from, and need not imply, the proportional win-fractions assumption
of \citet{MaoWang2021}. The latter requires the covariate-specific win ratio
to satisfy a time-invariant proportional structure, which may fail even when
separate PH models for the component event times provide reasonable working
models. Thus, in settings where the proportional win-fractions assumption is
questionable but the marginal PH assumptions remain plausible, the PH
model-based estimator provides a useful intermediate approach between the
more restrictive Mao--Wang model and the nonparametric local estimator. An
additional practical advantage is that the marginal PH assumptions can be
assessed using well-established diagnostic methods \citep{Schoenfeld1982,GrambschTherneau1994}.

We establish asymptotic properties of the log win ratio estimator on a fixed grid of the covariate
and its best linear projection, together with bootstrap inference and a
grid-based test of log-linearity. We also develop asymptotic theory and
bootstrap inference for the PH model-based estimator. Simulations evaluate
estimation accuracy, uncertainty quantification, robustness to model
misspecification, and performance of the log-linearity test. We illustrate
the methods using data from CALGB 9633 with death prioritized over disease recurrence, showing how covariate-specific win ratio curves and their low-dimensional summaries can characterize treatment-effect
heterogeneity across baseline characteristics such as age, tumor size, and performance status.

The remainder of the paper is organized as follows. Section~2 introduces the
notation, defines the covariate-specific win ratio, presents the local IPCW
and PH model-based estimators, and describes fixed-grid summaries, bandwidth
selection, and test for log-linearity. In this section, we also describe the bootstrap
procedures used for inference on the estimated curves, best linear
projections, and log-linearity assessments. Section~3 establishes asymptotic
properties of the proposed estimators and bootstrap procedures. Section~4
evaluates finite-sample performance through simulation studies. Section~5
illustrates the proposed methods using data from the lung cancer clinical
trial, and Section~6 concludes with a discussion.

%\vspace{-2.2em}
\section{Methods}

\subsection{Notation}

Let \(T_1\) and \(T_2\) denote the primary and secondary event times,
respectively, with \(T_1\) prioritized over \(T_2\), and let \(C\) be a
common right-censoring time. When \(T_1\) is a terminal event, such as death,
\(T_2\) denotes the underlying secondary event time and is observed only if
it occurs before \(T_1\). In this setting, define
\(
U_1=\min(T_1,C),\
\Delta_1=I(T_1\le C),
\)
\(
U_2=\min(T_2,T_1,C),\) and \(\Delta_2=I\{T_2\le \min(T_1,C)\}.
\)
Thus, secondary event follow-up terminates at the earlier of the terminal
event and censoring. When the primary event is nonterminal, the usual
definitions \(U_2=\min(T_2,C)\) and \(\Delta_2=I(T_2\le C)\) apply.

Let \(A\in\{0,1\}\) denote treatment and
\(X\in\mathcal X\subseteq\mathbb R^d\) baseline covariates. We assume conditional independent
censoring,
\(
C\perp(T_1,T_2)\mid(X,A).
\)
For subject \(i=1,\ldots,n\), the observed i.i.d. data are
\(
O_i=(U_{1i},\Delta_{1i},U_{2i},\Delta_{2i},X_i,A_i).
\)
Let
\(n_a\)
denote the number of subjects in treatment arm \(a\).

\vspace{-0.421em}
\subsection{Estimation and fixed-grid summaries of the covariate-specific win ratio}

For a fixed time horizon \(\tau\), define
\(
S_{1a}(t\mid x)
=
P(T_1>t\mid X=x,A=a),
\)
\(
J_a(\tau,u\mid x)
=
P(T_1>\tau,T_2>u\mid X=x,A=a),
\)
and
\(
G_a(c\mid x)
=
P(C\ge c\mid X=x,A=a).
\)
Consider two independent subjects with covariate value \(x\), one treated and one control, with prioritized event times \((T_{11},T_{21})\) and \((T_{10},T_{20})\), respectively. Suppose that longer event times are more favorable and that \(T_1\) has higher priority than \(T_2\). At the fixed horizon \(\tau\), the treated subject is said to \emph{win} if the treated subject has a longer time to the primary event than the control subject and the control primary event occurs by \(\tau\). If neither subject experiences the primary event by \(\tau\), the comparison proceeds to the secondary outcome, and the treated subject wins if the treated subject has a longer time to the secondary event and the control secondary event occurs by \(\tau\). Conversely, the treated subject is said to \emph{lose} if the control subject wins under the same prioritized comparison rule. Pairs that cannot be ordered by \(\tau\) contribute neither a win nor a loss.

Accordingly, the covariate-specific win and loss probabilities at horizon \(\tau\) are
\[
W(x;\tau)
=
P\{T_{11}>T_{10},T_{10}\le \tau\mid X=x\}
+
P\{T_{11}>\tau,T_{10}>\tau,T_{21}>T_{20},T_{20}\le\tau\mid X=x\},
\]
and
\[
L(x;\tau)
=
P\{T_{10}>T_{11},T_{11}\le \tau\mid X=x\}
+
P\{T_{10}>\tau,T_{11}>\tau,T_{20}>T_{21},T_{21}\le\tau\mid X=x\},
\]
respectively. Equivalently, they can be written as
\citep{Oakes2016,Parner2025},
\[
\begin{aligned}
W(x;\tau)
&=
-\int_0^\tau
S_{11}(t\mid x)\,dS_{10}(t\mid x)
-
\int_0^\tau
J_1(\tau,u\mid x)\,dJ_0(\tau,u\mid x),
\\
L(x;\tau)
&=
-\int_0^\tau
S_{10}(t\mid x)\,dS_{11}(t\mid x)
-
\int_0^\tau
J_0(\tau,u\mid x)\,dJ_1(\tau,u\mid x).
\end{aligned}
\]
The covariate-specific win ratio is defined as
\[
WR(x;\tau)=W(x;\tau)/L(x;\tau).
\]

Under conditional independent censoring, the survival components in the win
and loss probabilities satisfy
\[
S_{1a}(t\mid x)
=
E\left\{
\frac{I(U_1>t)}
     {G_a(t\mid X)}
\;\middle|\;
X=x,A=a
\right\},
\]
and
\[
J_a(\tau,u\mid x)
=
E\left\{
\frac{I(U_1>\tau,U_2>u)}
     {G_a(\tau\vee u\mid X)}
\;\middle|\;
X=x,A=a
\right\}.
\]
The factor \(G_a(\tau\vee u\mid X)\) arises because observing
\(\{T_1>\tau,T_2>u\}\) requires censoring-free follow-up through the larger
of \(\tau\) and \(u\), as in IPCW estimation of bivariate survival under
univariate censoring \citep{LinYing1993}. In the terminal event setting,
this event is identifiable because \(T_1>\tau\) ensures that the terminal
event has not occurred before the time interval relevant to the secondary
comparison.

Let \(K\) be a bounded symmetric second-order kernel and
\(
K_h(v)=h^{-d}K(v/h).
\)
For arm \(a\), define the normalized local weights
\[
\omega_{ia}(x)
=
K_h(X_i-x)I(A_i=a)\;\Big/\;
\sum_{j:A_j=a}K_h(X_j-x).
\]
We estimate the common conditional censoring survival function
\(G_a(c\mid x)\) by a local product-limit estimator
\(\widehat G_a(c\mid x)\) within treatment arm \(a\), using kernel weights
proportional to \(K_h(X_i-x)\)
\citep{Beran1981,Dabrowska1987}. In the terminal event setting, the
censoring distribution can be estimated from the primary event follow-up
\(
U_{1i}=\min(T_{1i},C_i)
\)
by treating
\(
1-\Delta_{1i}
\)
as the censoring event indicator. With no ties this is equivalent to
\(I(C_i<T_{1i})\). The local IPCW estimators of the marginal and joint survival functions are
\[
\widehat S_{1a}(t\mid x)
=
\sum_{i:A_i=a}
\omega_{ia}(x)
I(U_{1i}>t)
/\widehat G_a(t\mid x),
\]
and
\[
\widehat J_a(\tau,u\mid x)
=
\sum_{i:A_i=a}
\omega_{ia}(x)
I(U_{1i}>\tau,U_{2i}>u)
/\widehat G_a(\tau\vee u\mid x).
\]
Because the secondary contribution to the fixed-horizon win ratio involves
\(0\le u\le\tau\),
\(
\tau\vee u=\tau,
\)
and hence
\[
\widehat J_a(\tau,u\mid x)
=
\sum_{i:A_i=a}
\omega_{ia}(x)
\frac{
I(U_{1i}>\tau,U_{2i}>u)
}{
\widehat G_a(\tau\mid x)
},
\qquad 0\le u\le\tau.
\]
Because \(\widehat G_a(\cdot\mid x)\) is estimated locally at the target
covariate value \(x\), the censoring weight at a given time is common across
subjects in the local average for fixed \(x\), and subject-specific covariate
information enters through the kernel weights.

Let \(\widehat W_L(x;\tau)\) and \(\widehat L_L(x;\tau)\) be obtained by
substituting \(\widehat S_{1a}\) and \(\widehat J_a\) into the 
integral representations of \(W(x;\tau)\) and \(L(x;\tau)\). The local IPCW
estimator of the covariate-specific win ratio is
\[
\widehat{WR}_L(x;\tau)
=
\widehat W_L(x;\tau)
/
\widehat L_L(x;\tau).
\]

For descriptive low-dimensional summaries, the local estimator can be
evaluated on a fixed grid
\(
\mathcal G=\{x_1,\ldots,x_K\}
\)
of interior covariate values. Define
\(
\ell(x)=\log WR(x;\tau),
\
\widehat\ell_L(x)=\log\widehat{WR}_L(x;\tau),
\)
and
\[
\ell_{\mathcal G}
=
\{\ell(x_1),\ldots,\ell(x_K)\}^T,
\qquad
\widehat\ell_{L,\mathcal G}
=
\{\widehat\ell_L(x_1),\ldots,
  \widehat\ell_L(x_K)\}^T.
\]
To obtain interpretable low-dimensional summaries, we consider fixed-grid projections of \(\ell_{\mathcal G}\), viewed as descriptive approximations rather than model assumptions \citep{Chernozhukov2025}. Let \(\mathcal G=\{x_1,\ldots,x_K\}\) be a prespecified evaluation grid and let \(q(x)\) be a \(p\)-dimensional basis vector, for example \(q(x)=(1,x^T)^T\). Let \(D\) be the \(K\times p\) matrix whose \(k\)th row is \(q(x_k)^T\), and let \(Q\) be a positive diagonal matrix of prespecified grid weights. Assuming \(D^TQD\) is nonsingular, define the weighted best linear projection
\(
\beta^*=(D^TQD)^{-1}D^TQ\ell_{\mathcal G},
\)
with plug-in estimator
\(
\widehat\beta_L=(D^TQD)^{-1}D^TQ\widehat\ell_{\mathcal G}.
\)
The projection coefficients provide descriptive low-dimensional summaries of the covariate-specific log win ratio function rather than imposing a parametric assumption on its shape.

\subsection{Connection with estimators based on pairwise comparisons}

Traditional estimators of win ratio are based on pairwise comparisons between subjects in the treatment and subjects in the control group, including IPCW based pairwise comparison estimators  for
right-censored data \citep{Dong2020}. \citet{Parner2025} subsequently showed that an estimator obtained from the integral representation of the win and loss probabilities can also be written in
a pairwise comparison form, with censoring weights that appropriately reflect the
primary and secondary event comparisons. Our proposed local IPCW estimator admits an analogous representation.
The difference is that our target parameter is the covariate-specific win ratio, so
the corresponding pairwise comparison additionally incorporates local
kernel weights in the covariate space. Proposition~\ref{prop:pairwise}
establishes this equivalence, the proof of which is deferred to the Supplementary Material.

\begin{proposition}
\label{prop:pairwise}
Under the common right-censoring setup, suppose that
\(\widehat G_a(t\mid x)>0\) for \(0\le t\le\tau\), \(a=0,1\), where
\(\widehat G_a(\cdot\mid x)\) is the local product-limit estimator
constructed using the same local kernel weights as the survival estimators,
and assume that primary event and censoring times do not tie. Then the local
IPCW estimators of the win and loss probabilities have the exact representations
\[
\widehat W(x;\tau)
=
\sum_{i:A_i=1}\sum_{j:A_j=0}
\omega_{i1}(x)\omega_{j0}(x)
\phi_{ij}^{W}(x;\tau),
\]
and
\[
\widehat L(x;\tau)
=
\sum_{i:A_i=1}\sum_{j:A_j=0}
\omega_{i1}(x)\omega_{j0}(x)
\phi_{ij}^{L}(x;\tau),
\]
where
\[
\phi_{ij}^{W}(x;\tau)
=\frac{
\Delta_{1j} I(U_{1j}\le\tau)
I(U_{1i}>U_{1j})
}{
\widehat G_1(U_{1j}\mid x)
\widehat G_0(U_{1j}\mid x)
}+\frac{
I(U_{1i}>\tau,U_{1j}>\tau)
\Delta_{2j} I(U_{2j}\le\tau)
I(U_{2i}>U_{2j})
}{
\widehat G_1(\tau\mid x)
\widehat G_0(\tau\mid x)
},
\]
and
\[
\phi_{ij}^{L}(x;\tau)
=\frac{
\Delta_{1i} I(U_{1i}\le\tau)
I(U_{1j}>U_{1i})
}{
\widehat G_1(U_{1i}\mid x)
\widehat G_0(U_{1i}\mid x)
}
+\frac{
I(U_{1i}>\tau,U_{1j}>\tau)
\Delta_{2i} I(U_{2i}\le\tau)
I(U_{2j}>U_{2i})
}{
\widehat G_1(\tau\mid x)
\widehat G_0(\tau\mid x)
}.
\]
Thus, when the same local kernel weights and censoring estimator are used,
the integral estimator and the corresponding kernel-weighted IPCW pairwise comparison
estimator are  identical.
\end{proposition}

As an immediate special case, in the absence of censoring
\(\widehat G_a(t\mid x)\equiv1\), and the estimator reduces to the
conventional kernel-weighted pairwise win--loss estimator.

The proposition  may be viewed as a covariate-specific extension
of the  equivalence in \citet{Parner2025}. When the same
local kernel weights and censoring estimator are used, the integral and
kernel-weighted IPCW pairwise comparison estimators are algebraically identical and
should therefore be viewed as two representations of the same estimator
rather than as competing procedures. However, the integral
representation is particularly useful because it expresses the
covariate-specific win ratio directly in terms of conditional marginal and
joint survival functions, thereby facilitating alternative model-based
estimation and extensions to more complex censoring schemes, such as
interval censoring.

\subsection{Bandwidth selection}
\label{sec:bandwidth}

The bandwidth controls the bias--variance tradeoff of the local IPCW
estimator \citep{WandJones1995,FanGijbels1996}. In the asymptotic theory,
\(h\) is treated as deterministic and scalar for simplicity, whereas in
implementation with multivariate covariates we use a componentwise
bandwidth vector.

For pointwise estimation, we use a simple stability-based selector. Let
\(X=(X_1,\ldots,X_d)^T\), let \(N\) denote the pooled sample size
across treatment arms, and define the rule-of-thumb vector
\[
h_0
=
\left\{
\widehat{\mathrm{sd}}(X_1),\ldots,
\widehat{\mathrm{sd}}(X_d)
\right\}
N^{-1/(d+4)}.
\]
For a finite set \(\mathcal C\) of positive multipliers, let
\(
\mathcal H=\{c h_0:c\in\mathcal C\}.
\)
Let \(\mathcal B=\{z_1,\ldots,z_J\}\) denote a prespecified grid of
interior covariate values. For each \(h\in\mathcal H\), compute
\(\widehat\ell_h(z_j)=\log\{\widehat{WR}_h(z_j;\tau)\}\) on
\(\mathcal B\), and define
\[
Q(h)
=
\frac{1}{J}\sum_{j=1}^{J}
\left\{
\widehat\ell_h(z_j)
-
\widehat\ell_{c_{\rm ref}h}(z_j)
\right\}^{2},
\]
where \(c_{\rm ref}>1\) is fixed. If
\(Q_{\min}=\min_{h\in\mathcal H}Q(h)\), we select
\[
\widehat h
=
\min\left\{
h\in\mathcal H:
Q(h)\le \mathrm{tol}\,Q_{\min}
\right\},
\]
where the minimum is taken according to the ordering of
\(\mathcal C\) and \(\mathrm{tol}\ge1\). Thus, the selected bandwidth is
the smallest candidate for which the estimated log win ratio surface is
stable to modest additional smoothing.

For procedures based on best linear projections and for the log-linearity test, we instead use a rule-of-thumb bandwidth computed from the original pooled sample. In the one-dimensional setting, we use the normal-reference bandwidth
\(
h=1.06,\widehat{\mathrm{sd}}(X)N^{-1/5},
\)
following the classical normal-reference rule \citep{Silverman1986}. In the multivariate setting, for simplicity and to maintain a comparable smoothing scale across covariates, we use
\(
h_j=\widehat{\mathrm{sd}}(X_j)N^{-1/5},\ j=1,\ldots,d.
\)

\subsection{An estimator based on proportional hazards model and copula dependence}
\label{sec:model-based}

The integral representation of the win ratio suggests  a natural
model-based estimator obtained by specifying regression models for the
marginal event-time distributions and a copula model for their dependence.
Unlike the proportional win-fractions model, this approach does not impose a
log-linear or proportional structure directly on the win ratio. Instead,
the covariate-specific win ratio is induced by the fitted joint distribution
of the prioritized outcomes and may therefore vary nonlinearly with the
covariates.

For treatment arm \(a\in\{0,1\}\) and outcome \(j\in\{1,2\}\), assume a
proportional hazards model
\(
\lambda_{ja}(t\mid x)
=
\lambda_{ja,0}(t)\exp(x^T\beta_{ja}),
\)
where \(\lambda_{ja,0}(t)\) is an unspecified baseline hazard and
\(\beta_{ja}\) is a regression coefficient vector. Let
\(
\Lambda_{ja,0}(t)
\)
be the baseline cumulative hazard function. The conditional survival function is
\(
S_{ja}(t\mid x)
=
\exp\left\{
-\Lambda_{ja,0}(t)\exp(x^T\beta_{ja})
\right\},
\)
with density denoted as
\(
f_{ja}(t\mid x).
\)

Dependence between \(T_1\) and \(T_2\) within treatment arm \(a\) is
represented through a survival-scale copula \(C_a\)  \citep{Nelsen2006}. Specifically,
\[
J_a(s,t\mid x)
=
P(T_1>s,T_2>t\mid X=x,A=a)
=
C_a\left\{
S_{1a}(s\mid x),S_{2a}(t\mid x);
\gamma_a
\right\},
\]
where \(\gamma_a\) denotes the copula parameter vector. The construction is not tied to a particular copula family. In the simulation studies, the copula family was chosen according to the
data-generating mechanism, whereas in the C9633 application we used
the two-parameter BB1 family, which allows flexible asymmetric tail
dependence.

For flexible estimation of the marginal baseline hazards, we use a Bernstein
polynomial sieve approach \citep{ZhouHuSun2017,SunDing2021}. Let \(0\le c<u\) cover the follow-up interval and define
\(
s(t)=(t-c)/(u-c).
\)
For sieve degree \(m\), let
\[
B_k(t;m,c,u)
=
{m\choose k}s(t)^k\{1-s(t)\}^{m-k},
\qquad k=0,\ldots,m.
\]
The baseline cumulative hazard function in arm \(a\) for outcome \(j\) is approximated
by
\[
\Lambda_{ja,0}(t)
=
\sum_{k=0}^m \phi_{jak} B_k(t;m,c,u),
\]
subject to constraint
\(
0=\phi_{ja0}\le \phi_{ja1}\le\cdots\le\phi_{jam},
\)
which guarantees monotonicity of the estimated cumulative baseline hazard.
A convenient unconstrained parameterization is
\[
\phi_{ja0}=0,
\qquad
\phi_{jak}
=
\sum_{r=1}^k \exp(\psi_{jar}),
\qquad k=1,\ldots,m.
\]

For treatment arm \(a\), let \(\Theta_{a,m}\) denote the corresponding
Bernstein sieve parameter space for
\(
\theta_a
=
(\beta_{1a},\beta_{2a},\gamma_a,\Lambda_{1a},\Lambda_{2a})
\),
with each baseline cumulative hazard represented by the degree-\(m\)
Bernstein expansion above and satisfying the stated monotonicity
constraints. The regression coefficients, copula parameter, and sieve
coefficients are estimated separately within each treatment arm by
maximizing the likelihood function over \(\Theta_{a,m}\).

For subject \(i\), let \(Y_{1i}=U_{1i}\) and \(Y_{2i}=U_{2i}\), with
\(\Delta_{1i}\) and \(\Delta_{2i}\) defined in Section~2.1. Thus, in the
terminal event setting,
\(
Y_{1i}=\min(T_{1i},C_i),
\
Y_{2i}=\min(T_{2i},T_{1i},C_i).
\)
For a subject in treatment arm \(a\), write
\(S_{ji}=S_{ja}(Y_{ji}\mid X_i)\), \(j=1,2\), and define
\[
C_a^{(r,s)}(u,v)
=
\frac{\partial^{r+s}C_a(u,v)}
{\partial u^r\partial v^s},
\qquad
C_a^{(0,0)}=C_a.
\]
Up to factors involving only the censoring distribution, the individual  likelihood contribution is
\[
L_{ia}
=
\left\{f_{1a}(Y_{1i}\mid X_i)\right\}^{\Delta_{1i}}
\left\{f_{2a}(Y_{2i}\mid X_i)\right\}^{\Delta_{2i}}
C_a^{(\Delta_{1i},\Delta_{2i})}(S_{1i},S_{2i}).
\]
When the terminal event occurs before the secondary event,
\(\Delta_{1i}=1\), \(\Delta_{2i}=0\), and \(Y_{2i}=Y_{1i}\), so this
likelihood contribution integrates over the unobserved region
\(T_{2i}>T_{1i}\) rather than treating the terminal event as independent
censoring of \(T_{2i}\). Let \(\widehat\theta_a\) denote the resulting sieve maximum likelihood estimator, and let \(\widehat S_{ja}^{\,M}\) and
\(\widehat J_a^{\,M}\) denote the fitted marginal and joint survival
functions.

The PH model-based win and loss probabilities are obtained by substituting
these fitted survival functions into the same fixed-horizon representations
used to define the local estimator:
\[
\widehat W_M(x;\tau)
=-\int_0^\tau
\widehat S_{11}^{\,M}(t\mid x)
\,d\widehat S_{10}^{\,M}(t\mid x)
-\int_0^\tau
\widehat J_1^{\,M}(\tau,t\mid x)
\,d\widehat J_0^{\,M}(\tau,t\mid x),
\]
and
\[
\widehat L_M(x;\tau)
=-\int_0^\tau
\widehat S_{10}^{\,M}(t\mid x)
\,d\widehat S_{11}^{\,M}(t\mid x)
-\int_0^\tau
\widehat J_0^{\,M}(\tau,t\mid x)
\,d\widehat J_1^{\,M}(\tau,t\mid x).
\]
The corresponding model-based covariate-specific win ratio is
\[
\widehat{WR}_M(x;\tau)
=
\widehat W_M(x;\tau)/\widehat L_M(x;\tau).
\]

This construction differs fundamentally from imposing proportional win
fractions. Although proportional hazards models are used for the component
event-time distributions, \(WR_M(x;\tau)\) is a nonlinear functional of the
two marginal survival distributions and their dependence structure.
Consequently, neither \(\log WR_M(x;\tau)\) nor its covariate effect is
required to be constant or linear in \(x\). The model-based estimator can
therefore accommodate covariate-specific patterns in the win ratio while
potentially gaining efficiency relative to the fully local estimator when
the marginal and dependence models are adequately specified. This is an attractive alternative to existing model-based win ratio estimation  because the PH assumption is widely adopted and plausible in many applications.

\subsection{Test for log-linearity of the win ratio function}
\label{sec:test_linearity}

The covariate-specific win ratio curve provides a flexible description of
treatment-effect heterogeneity, but a linear representation of the log win
ratio may offer a useful and more interpretable summary of the covariate effect. We therefore assess whether \(\ell(x)=\log\{\mathrm{WR}(x;\tau)\}\) can be adequately represented by a
linear function of the covariates. Tests are constructed from both the local
IPCW and PH copula estimators.

For the log-linearity test, we use the same weighted projection construction as in Section~2.2, specialized to the linear basis \(q(x)=(1,x^T)^T\). Let
\(
\mathcal G_n=\{x_{1n},\ldots,x_{K_n n}\}
\)
denote the evaluation grid, and let
\(
\ell_{\mathcal G_n}
=
\{\ell(x_{1n}),\ldots,\ell(x_{K_n n})\}^T.
\)
Unlike the fixed grid used for descriptive projection summaries, here we allow the grid size \(K_n\) and the grid points to depend on \(n\), so that the grid may become increasingly dense over the covariate region of interest, and a fixed grid is included as a special case.

Let \(D_n\) be the \(K_n\times(d+1)\) design matrix with \(k\)th row \((1,x_{kn}^T)\), and let
\(
Q_n=\operatorname{diag}\),
\(q_{kn}>0\), \(\sum_{k=1}^{K_n}q_{kn}=1.
\)
The corresponding weighted projection and residual operators are
\(
P_n=D_n(D_n^TQ_nD_n)^{-1}\) \(D_n^TQ_n,\) and
\(M_n=I_{K_n}-P_n.
\)
The null hypothesis of log-linearity can then be written as
\(
H_0:M_n\ell_{\mathcal G_n}=0,
\)
or equivalently \(\ell_{\mathcal G_n}\in\mathcal M_n\), where \(\mathcal M_n\) is the space of linear functions evaluated on \(\mathcal G_n\).

For the local IPCW estimator, let
\(\widehat\ell_{L,\mathcal G_n}\) denote the estimated log win ratio vector.
We use the lack-of-fit statistic
\[T_{L,n}=nh^d(M_n\widehat\ell_{L,\mathcal G_n})^T
Q(M_n\widehat\ell_{L,\mathcal G_n}).\]
Large values indicate departure from linearity. Because its null distribution
depends on kernel smoothing, censoring estimation, and the covariance
structure of the local estimator, critical values and \(p\) values are
obtained from the centered nonparametric bootstrap described in
Section~\ref{sec:bootstrap_inference}.

For the PH copula estimator, let \(\widehat\ell_{M,\mathcal G}\) denote the fitted log win ratio vector on a prespecified fixed grid \(\mathcal G\), with corresponding projection and residual matrices \(P\) and \(M=I-P\). In contrast to the local IPCW test, for which the evaluation grid may depend on \(n\) in the asymptotic formulation, the model-based test is defined on a fixed finite grid. The model-based lack-of-fit statistic is
\[
T_{M,n}
=
n(M\widehat\ell_{M,\mathcal G})^TQ
(M\widehat\ell_{M,\mathcal G}).
\]
Its null distribution is approximated using the centered weighted multiplier bootstrap of Section~\ref{sec:bootstrap_inference}; in each bootstrap sample, the PH--copula model is refitted and the statistic is computed from
\(
M(\widehat\ell_{M,\mathcal G}^*-\widehat\ell_{M,\mathcal G}).
\)
The PH copula model-based test can be more powerful when the working marginal and copula models are adequate, whereas the local IPCW test relies on substantially weaker structural assumptions.

When a linear representation is adequate, the components of
\(\widehat\beta\) provide a simple interpretation: a one-unit increase in the
\(j\)th covariate, holding the remaining covariates fixed, corresponds to an
approximately \(\exp(\widehat\beta_j)\) fold change in the win ratio. Failure
to reject linearity should not, however, be interpreted as evidence that the
true function is exactly linear, since moderate nonlinear departures may be
difficult to detect in finite samples.

The projection coefficients remain meaningful even when linearity fails,
because they represent the best linear projection of the covariate-specific
log win ratio under the chosen weighting measure. Thus, when linearity is not
rejected, the projection may provide a parsimonious summary; when it is
rejected, the projection still summarizes the overall linear trend, but the
full covariate-specific curve is needed to describe nonlinear features. This
interpretation is consistent with the use of best linear projections under
model misspecification in the heterogeneous treatment effect literature \citep[e.g.][]{Chernozhukov2025}.

\subsection{Bootstrap inference}
\label{sec:bootstrap_inference}

Direct analytic variance estimation is cumbersome because the estimators involve estimated nuisance quantities and nonlinear functionals of the observed data, with additional complexity arising from censoring adjustment and, for the local IPCW estimator, kernel smoothing.  We therefore use bootstrap procedures for inference on the covariate-specific win ratio curve and derived summaries. Because the local IPCW and PH copula based estimators have different asymptotic structures, their bootstrap implementations are described separately below.

For the local IPCW estimator, the limiting covariance structure depends on
kernel smoothing, censoring estimation, and nonlinear plug-in functionals of
the marginal and joint survival curves. We therefore use a nonparametric
bootstrap in which subjects are resampled with replacement within treatment
arms. In each bootstrap replication, the local IPCW estimator is recomputed
on the same evaluation grid
\(\mathcal G_n=\{x_1,\ldots,x_{K_n}\}\), and all derived quantities are
recalculated. Let
\(
\widehat\ell_{L,\mathcal G_n}^{\,*}
=
\{\widehat\ell_L^{\,*}(x_1),\ldots,
\widehat\ell_L^{\,*}(x_{K_n})\}^T
\)
denote the bootstrap estimate, where
\(\widehat\ell_L(x)=\log\widehat{WR}_L(x;\tau)\).

For fixed-grid inference, the empirical distribution of
\(\widehat\ell_{L,\mathcal G_n}^{\,*}
-\widehat\ell_{L,\mathcal G_n}\)
is used to approximate the sampling distribution of
\(\widehat\ell_{L,\mathcal G_n}-\ell_{\mathcal G_n}\).
Pointwise confidence intervals for \(WR(x_k;\tau)\) are obtained by forming
bootstrap intervals for \(\ell(x_k)\) and exponentiating. The same bootstrap
samples are used for inference on the best linear projection and other
fixed-dimensional summaries. For example, if
\(\widehat\gamma_L=A\widehat\ell_{L,\mathcal G_n}\), then
\(\widehat\gamma_L^{\,*}=A\widehat\ell_{L,\mathcal G_n}^{\,*}\), and the
empirical distribution of
\(\widehat\gamma_L^{\,*}-\widehat\gamma_L\) is used for inference.

For the log-linearity test, we use the centered bootstrap statistic
\[
T_{L,n}^*
=
nh^d
\left\{
M_n
\left(
\widehat\ell_{L,\mathcal G_n}^{\,*}
-
\widehat\ell_{L,\mathcal G_n}
\right)
\right\}^T
Q
\left\{
M_n
\left(
\widehat\ell_{L,\mathcal G_n}^{\,*}
-
\widehat\ell_{L,\mathcal G_n}
\right)
\right\},
\]
where \(M_n=I_{K_n}-P_n\) is the residual projection matrix defined in
Section~\ref{sec:test_linearity}. The bootstrap critical value is obtained
from the appropriate upper quantile of the bootstrap distribution of
\(T_{L,n}^*\), and the bootstrap \(p\) value is
\(
p^*
=
B^{-1}
\sum_{b=1}^B
I\{T_{L,n}^{*(b)}\ge T_{L,n}\}.
\)
The bandwidth is selected once from the original data and held fixed across bootstrap resamples.

For the PH copula estimator, no kernel smoothing or bandwidth selection is
involved. We use an arm-specific weighted multiplier bootstrap, which preserves all observations in each treatment arm  and is convenient for the PH copula sieve likelihood \citep{PraestgaardWellner1993}. Independently of the observed data, let \(\xi_{ai}\), for subjects with \(A_i=a\), be
independent standard exponential random variables with mean one and variance
one. For treatment arm \(a\), the bootstrap empirical criterion is obtained
by replacing the ordinary empirical average by
\[
\mathbb P_{n,a}^*f
=
\frac{1}{n_a}
\sum_{i:A_i=a}\xi_{ai}f(O_i).
\]
The PH and copula model is refitted by maximizing the corresponding weighted
likelihood using the same sieve dimension \(m_n\) as in the original fit.
The resulting estimator is denoted by \(\widehat\theta_{a,m_n}^*\), and
\(\widehat{WR}_M^*(x;\tau)\) is recomputed on the same fixed evaluation grid
\(\mathcal G=\{x_1,\ldots,x_K\}\).

Let
\[
\widehat{\boldsymbol\ell}_{M,\mathcal G}^{\,*}
=
\left\{
\log\widehat{WR}_M^{\,*}(x_1;\tau),\ldots,
\log\widehat{WR}_M^{\,*}(x_K;\tau)
\right\}^T.
\]
For pointwise and fixed-grid inference, the conditional multiplier bootstrap
distribution of
\(
\sqrt n(\widehat{\boldsymbol\ell}_{M,\mathcal G}^{\,*}
-\) \(
\widehat{\boldsymbol\ell}_{M,\mathcal G})
\)
is used to approximate the sampling distribution of
\(
\sqrt n
\left(
\widehat{\boldsymbol\ell}_{M,\mathcal G}
-
\boldsymbol\ell_{\mathcal G}
\right).
\)
Pointwise confidence intervals and intervals for fixed-dimensional summaries,
including the coefficients of the best linear projection, are obtained from
the corresponding multiplier-bootstrap distributions. In particular, the
best linear projection is recomputed from each bootstrap win ratio curve.

For the log-linearity test, we use the centered multiplier-bootstrap statistic
\[T_{M,n}^*
=
n\{M(\widehat{\boldsymbol\ell}_{M,\mathcal G}^{\,*}
-\widehat{\boldsymbol\ell}_{M,\mathcal G})\}^T
Q
\{M(\widehat{\boldsymbol\ell}_{M,\mathcal G}^{\,*}
-\widehat{\boldsymbol\ell}_{M,\mathcal G}),\]\
where \(M=I-P\) is the residual projection matrix on the fixed grid.
The bootstrap critical value is the appropriate upper quantile of the
conditional distribution of \(T_{M,n}^*\), and the corresponding bootstrap
\(p\) value is
\(B^{-1}\sum_{b=1}^B I\{T_{M,n}^{*(b)}\geq T_{M,n}\}\).

\subsection{Software implementation}
\label{sec:software}

An R package implementing the proposed methods is available at
\texttt{https://github.com/zhiguoli9 19/winratioReg}. The package provides estimation of the fixed-horizon covariate-specific win ratio using both the local IPCW and
PH copula model-based methods, pointwise bootstrap inference for the estimated win
ratio curve, best linear projection summaries with bootstrap standard errors
and confidence intervals, and the corresponding tests of log-linearity.
The PH copula implementation supports multiple copula families and reports
maximized log likelihood, AIC, and BIC to facilitate comparison among
candidate dependence models. It also reports estimated copula parameters and
associated dependence summaries, allowing sensitivity analyses across copula
specifications. Functions for evaluating and plotting the estimated
covariate-specific win ratio curves and their best linear projections are
also provided. For the local IPCW procedure, bandwidth selection is performed
automatically using the data-adaptive procedure described in
Section~\ref{sec:bandwidth}, with the option for users to specify
bandwidths directly.

\section{Asymptotic results}

We present asymptotic results separately for the local IPCW estimator and
the PH model-based estimator. The two estimators have different rates and
sources of uncertainty. The local estimator involves kernel smoothing and
converges at the nonparametric rate \((nh^d)^{-1/2}\), whereas the
model-based estimator is a smooth functional of the fitted PH and copula
model and, under correct model specification and suitable sieve regularity
conditions, has \(\sqrt{n}\) rate of convergence. These results also justify the
corresponding bootstrap procedures for fixed-dimensional summaries and
the log-linearity tests. Assumptions and proofs are provided in the
Supplementary Material.

\subsection{Local IPCW estimator}

We state asymptotic results for the local IPCW estimator on an evaluation grid \(\mathcal G_n=\{x_{1n},\ldots,x_{K_n n}\}\) contained in a compact interior region of the covariate support. The case \(K_n=K\) gives fixed-grid inference for \(\widehat\ell_{\mathcal G}\) and projection summaries, whereas \(K_n\to\infty\) supports grid-based lack-of-fit testing. 

\begin{theorem}
\label{thm:master_grid}
Suppose that Assumptions (A1)--(A5) hold uniformly over \(\mathcal G_n\), and that \(n_a/n\to\pi_a\in(0,1)\), \(a=0,1\). Then, for each \(x\in\mathcal G_n\),
\[
\widehat\ell(x)-\ell(x)
=
h^2B_\ell(x)+Z_n(x)+R_n(x),
\]
where \(B_\ell(x)\) is the leading second-order smoothing bias, \(Z_n(x)\) is the leading kernel-weighted influence function term, and
\[
\max_{x\in\mathcal G_n}|R_n(x)|=o_p\{(nh^d)^{-1/2}\}.
\]
Equivalently,
\[
\sqrt{nh^d}
\{\widehat\ell(x)-\ell(x)-h^2B_\ell(x)\}
=
\sqrt{nh^d}\,Z_n(x)+o_p(1)
\]
uniformly over \(x\in\mathcal G_n\). For any fixed finite points in \(\mathcal G_n\), the corresponding vector of \(\sqrt{nh^d}Z_n(\cdot)\) is asymptotically
multivariate normal.
\end{theorem}

\begin{corollary}
\label{cor:grid_asymp}
Suppose that the conditions of Theorem~\ref{thm:master_grid} hold with \(\mathcal G_n=\mathcal G=\{x_1,\ldots,x_K\}\), where \(K\) is fixed. Then
\[
\sqrt{nh^d}
(\widehat\ell_{\mathcal G}
-\ell_{\mathcal G}
-b_h)
\overset{d}\to
N(0,\Sigma_{\ell}),
\]
where \(b_h=h^2\{B_{\ell}(x_1),\ldots,B_{\ell}(x_K)\}^T\). If we further assume that \(nh^{d+4}\to0\), then
\[
\sqrt{nh^d}
(\widehat\ell_{\mathcal G}
-\ell_{\mathcal G})
\overset{d}\to
N(0,\Sigma_{\ell}).
\]
\end{corollary}

\begin{corollary}
\label{cor:linear_projection_asymp}
Let \(\widehat\beta\), \(\beta^*\), and \(A\) be defined as in Section~2.2. Under the conditions of Corollary~\ref{cor:grid_asymp},
\[
\sqrt{nh^d}
(\widehat\beta-\beta^*-A b_h)
\overset{d}\to
N_p(0,A\Sigma_{\ell}A^T).
\]
If we further have \(nh^{d+4}\to0\), then
\[
\sqrt{nh^d}
(\widehat\beta-\beta^*)
\overset{d}\to
N_p(0,A\Sigma_{\ell}A^T).
\]
\end{corollary}

Theorem~\ref{thm:grid_integrated_bootstrap} below establishes bootstrap validity
for the grid-based test of log-linearity. Here $k_n\to\infty$ and it satisfies the rate condition in Assumption~A(5).  

\begin{theorem}
\label{thm:grid_integrated_bootstrap}
Suppose the conditions of Theorem~\ref{thm:master_grid} hold,
\(nh^{d+4}\to0\), and Assumption~(A6) holds for the selected growing grid.
Then, under \(H_0:\ell_{\mathcal G_n}\in\mathcal M\),
\[
\sup_{t\in\mathbb R}
\left|
P^*(T_{Ln}^*\le t)
-
P(T_{Ln}\le t)
\right|
\overset{p}\to0.
\]
\end{theorem}

\subsection{PH Model-based estimator}

Let \(\widehat\theta_{a,m_n}\) denote the sieve maximum likelihood estimator
of the PH and copula model in treatment arm \(a\), and let
\(
\widehat\ell_M(x)
=
\log\{\widehat{WR}_M(x;\tau)\}
\)
denote the corresponding model-based estimator of
\(\ell(x)=\log\{WR(x;\tau)\}\). Under correct specification of  PH models and copula dependence, appropriate
regularity conditions for the sieve likelihood, and sufficiently small sieve
approximation bias, the win ratio estimator is asymptotically linear at the
\(\sqrt n\)-rate, as stated in the following theorem.

\begin{theorem}
\label{thm:model_based_pointwise}
Suppose that Assumptions (M1)--(M6) hold. Then 
\[
\sqrt n
\{\widehat\ell_M(x)-\ell(x)\}
=
\frac{1}{\sqrt n}
\sum_{i=1}^n
\varphi_M(O_i;x)
+
o_p(1),
\]
for each $x\in\mathcal X$, where \(\varphi_M(O;x)\) is a mean-zero influence function. Consequently,
\[
\sqrt n
\{\widehat\ell_M(x)-\ell(x)\}
\overset{d}\longrightarrow
N\{0,\sigma_M^2(x)\},
\]
where
\(
\sigma_M^2(x)
=
E\{\varphi_M(O;x)^2\}.
\)
\end{theorem}

\begin{corollary}
\label{cor:model_based_grid}
Let
\(
\mathcal G=\{x_1,\ldots,x_K\}
\)
be a fixed evaluation grid and define
\(
\widehat\ell_{M,\mathcal G}
=
\{\widehat\ell_M(x_1),\ldots,
  \widehat\ell_M(x_K)\}^T.
\)
Under the conditions of
Theorem~\ref{thm:model_based_pointwise},
\[
\sqrt n
\left(
\widehat\ell_{M,\mathcal G}
-
\ell_{\mathcal G}
\right)
\overset{d}\longrightarrow
N(0,\Sigma_M),
\]
where the \((j,k)\)th element of \(\Sigma_M\) is
\(
E\{
\varphi_M(O;x_j)\varphi_M(O;x_k)
\}.
\)
Hence, for any fixed matrix \(A\),
\[
\sqrt n
A
\left(
\widehat\ell_{M,\mathcal G}
-
\ell_{\mathcal G}
\right)
\overset{d}\longrightarrow
N(0,A\Sigma_MA^T).
\]
\end{corollary}

By the following theorem, the weighted multiplier bootstrap provides
asymptotically valid pointwise and fixed-grid inference for the PH
model-based estimator.

\begin{theorem}
\label{thm:model_based_bootstrap}
Suppose that Assumptions~(M1)--(M6) hold. Let
\(\widehat\ell_{M,\mathcal G}^{\,*}\) denote the weighted multiplier
bootstrap analogue of \(\widehat\ell_{M,\mathcal G}\). Then, conditionally on the observed data,
\[
\sqrt n
\left(
\widehat\ell_{M,\mathcal G}^{\,*}
-
\widehat\ell_{M,\mathcal G}
\right)
\overset{d^*}\longrightarrow
N_K(0,\Sigma_M)
\]
in probability.
\end{theorem}

We next consider bootstrap validity for the PH copula model-based lack-of-fit test stated in the following corollary.

\begin{corollary}
\label{cor:model_based_linearity_test}
Under the conditions of Theorem~\ref{thm:model_based_bootstrap}, suppose
\(H_0:\ell_{\mathcal G}\in\mathcal M\), and let \(M=I-P\) be the residual
projection matrix defined in Section~\ref{sec:test_linearity}. Assume further
that \(M\Sigma_M M^T\neq 0\). Then the weighted multiplier bootstrap
consistently estimates the null distribution of the PH copula model-based
lack-of-fit statistic
\(T_{M,n}\).
In particular, 
\[
\sup_{t\in\mathbb R}
\left|
P^*(T_{M,n}^*\le t)
-
P(T_{M,n}\le t)
\right|
\overset{p}\longrightarrow0.
\]
\end{corollary}

\section{Simulation}

We conducted simulation studies to evaluate the finite-sample performance
of the proposed estimation and log-linearity testing procedures for
covariate-specific win ratios. For estimation, we compared the local IPCW
estimator with the regression method of \citet{MaoWang2021}, referred to
as the Mao--Wang method below, and with the proposed model-based
PH copula estimator. Particular attention was given to settings in which
the proportional win-fraction assumption underlying the Mao--Wang method
holds or fails, and in which the PH copula model is correctly specified
or misspecified.

\subsection{Data-generating mechanisms}

We focused on the more challenging setting in which the primary event is
terminal. The simulation study was designed to distinguish the structural
assumptions underlying the Mao--Wang proportional win-fractions estimator
from those underlying the model-based PH copula estimator. We considered
three classes of data-generating mechanisms for the estimation comparisons.

In the first class, the terminal and latent nonterminal event times were
generated from the Gumbel--Hougaard/Lehmann construction used by
\citet{MaoWang2021}. Covariates affect the two event-time margins through
proportional hazards models with common covariate effects. Under this
construction, the proportional win-fractions assumption holds and the
covariate-specific log win ratio is
\(
\log\{WR(x)\}=\log(1.5)+0.5x.
\)
At the same time, both marginal event time models satisfy proportional
hazards and the dependence structure is Gumbel--Hougaard. Thus, when the
corresponding Gumbel--Hougaard copula is used in the model-based analysis,
both the Mao--Wang and PH copula models are correctly specified.

In the second class, the terminal and latent nonterminal event times were
generated from proportional hazards models separately for the two endpoints
within each treatment arm, with exponential baseline hazards and dependence
induced by a Clayton survival copula. The treatment and covariate effects
were chosen so that the proportional win-fractions assumption was violated,
whereas the marginal proportional hazards models and Clayton copula were
correctly specified. This class therefore evaluates the two structured
estimators when the proportional win-fractions assumption fails but the
component event time and dependence models underlying the PH copula
estimator are correctly specified.

In the third class, we retained the Clayton survival copula but deliberately
misspecified the marginal proportional hazards models. The marginal hazards
were piecewise exponential, with covariate effects allowed to change after
prespecified time points. In addition, the hazard of the primary event included a
persistent treatment specific quadratic covariate effect. More specifically,
the hazard of the primary event was generated from a model of the form
\(
\lambda_{1a}(t\mid x)
=
\lambda_{1a,0}(t)
\exp\{\beta_{1a}(t)x-a x^2\},
\)
where \(\beta_{1a}(t)\) changed between early and late follow-up. The hazard of the secondary event also had time-varying covariate effects, but no quadratic term. Consequently, the working PH models with
time-invariant linear covariate effects were misspecified even though the
Clayton dependence model remained correctly specified. The proportional
win-fractions assumption was also violated. This class was designed as a
stress test setting in which the assumptions in both model-based approaches are violated, whereas
the local IPCW estimator does not rely on either structural assumption.

For all three classes for pointwise estimation, the continuous covariate \(X\) was generated from a
standard normal distribution truncated to \([-1,1]\). The covariate-specific
win ratio was evaluated at
\(
x\in\{-0.75,-0.375,0,0.375,0.75\}
\)
at the fixed horizon \(\tau=5\). Each treatment arm contained 250 subjects.
A random censoring time \(C^R\) was generated independently from an exponential
distribution with rate \(0.20\) and administratively truncated at
\(t_{\max}=8\). Thus,
\(
C=\min(C^R,8),
\)
and, for the terminal event setting,
\(
U_1=\min(T_1,C),\) and 
\(
U_2=\min(T_2,T_1,C),
\)
with
\(
\Delta_1=I(T_1\le C),\) and
\(
\Delta_2=I\{T_2\le\min(T_1,C)\}.
\)

The dependence strength in Classes~2 and~3 was chosen to correspond to
Kendall's \(\tau=0.30\). In Class~1, the population covariate-specific win
ratio is known from the data-generating construction. In Classes~2 and~3,
the population win ratio at \(\tau=5\) was obtained by numerical integration
and treated as the simulation truth.

We conducted a separate set of simulations to evaluate estimation of low-dimensional linear summaries of the covariate-specific log win ratio. These simulations involved two continuous covariates, \(X_1\) and \(X_2\), generated jointly with correlation \(0.3\), with 250 subjects per treatment arm and analysis horizon \(\tau=8\). We considered both exact and approximate low-dimensional structures. Exact constant or log-linear win ratio surfaces arise naturally under correctly specified proportional win-fractions models and therefore provide useful benchmark settings. In contrast, under data-generating mechanisms that violate the proportional win-fractions structure, it is difficult to construct realistic joint event time distributions for which the fixed-horizon win ratio remains exactly constant or exactly log-linear. We therefore also considered approximately constant and approximately log-linear settings, which reflect the practical situation in which a simple linear projection provides a useful summary even though the underlying win-ratio surface is not exactly linear.

Four scenarios were considered. In the first, the win ratio was constant in the covariates and the proportional win-fractions assumption held. In the second, the fixed-horizon win ratio was approximately constant, while the proportional win-fractions assumption was violated. The third scenario had an exactly log-linear win ratio under a correctly specified proportional win-fractions model, whereas in the fourth the  log win ratio was approximately linear while the proportional win-fractions assumption failed. In the approximately constant and approximately log-linear scenarios, the target coefficients were defined as the coefficients of the population linear projection of \(\log WR(X;\tau)\). These four settings were used to compare the Mao--Wang, PH copula model-based, and local IPCW estimators of the intercept and covariate slopes.

For the testing simulations, we considered both one- and two-dimensional
continuous covariate settings. The log-linearity test was evaluated at sample
sizes of 300, 500, and 800 subjects per treatment arm. In the one-dimensional
setting, the test was evaluated on 30 equally spaced grid points over
\([-1,1]\), whereas in the two-dimensional setting a \(5\times5\) grid over
\([-1,1]^2\) was used. The analysis horizon was \(\tau=5\), with
administrative censoring at \(t_{\max}=8\).

The null data generating mechanisms were calibrated so that the log win ratio function was approximately linear over the evaluation region.
An exact linear null is also difficult to construct in the present setting because
\(WR(x;\tau)\) is an induced nonlinear functional of the two marginal
event time distributions and their dependence structure, rather than a
directly parameterized regression function. Imposing exact linearity would
require solving nonlinear restrictions on the joint event time model over the
entire covariate region and could lead to highly specialized data-generating
mechanisms that are less representative of the general settings considered
here. We therefore used parameter configurations for which the population log
win ratio surface was numerically calibrated to be very close to a linear
function over the evaluation region. Accordingly, rejection rates under these
scenarios should be interpreted as near-null rejection rates rather than exact
type I error probabilities. Two alternative configurations were considered,
representing moderate and stronger departures from linearity.

\subsection{Simulation implementation}

Each estimation setting was repeated over 500 Monte Carlo replications.
Standard errors for the local IPCW and PH copula estimators were obtained
using 200 bootstrap samples within each replication, with
95\% confidence intervals based on normal approximations. For each estimation target,
performance was summarized by empirical bias, empirical standard error,
average estimated standard error, and empirical coverage probability of the 95\% confidence interval.
For the log-linearity test, we reported empirical rejection rates under the
near-null settings and under two alternatives representing moderate and
stronger departures from linearity. 

Bandwidths were selected using the procedures described in
Section~\ref{sec:bandwidth}. For the pointwise estimation, we
used the stability-based selector with tol=1.10, and for the best linear projection
and log-linearity test we used the corresponding rule-of-thumb bandwidth.
In all cases, the bandwidth was computed from the original sample and held
fixed throughout the bootstrap.

All simulation studies were implemented in R \citep{RCoreTeam2026}. The  code used to reproduce the simulation studies is available in the associated GitHub repository.

\subsection{Results}

Simulation results for estimation are presented in Tables~1--3 and
Figure~1. For notational simplicity, we omit \(\tau\) from
\(WR(X;\tau)\) throughout this section and write \(WR(X)\).
When the proportional win-fractions assumption held and the PH copula model was
correctly specified, both model-based estimators performed well. The
Mao--Wang estimator was generally the most efficient, the PH copula
estimator had comparable performance, and the local IPCW estimator showed
greater variability because of its nonparametric flexibility,
while remaining approximately unbiased with coverage close to the nominal
level. When the proportional win-fractions assumption was violated but the
PH copula model remained correctly specified, the Mao--Wang estimator
exhibited appreciable bias and corresponding undercoverage, whereas the
PH copula estimator remained approximately unbiased and was more efficient
than the local IPCW estimator. In the setting in which both the proportional
win-fractions assumption and the marginal PH models were misspecified, both
model-based estimators showed bias and under coverage, while the local IPCW
estimator remained stable, with small bias and near-nominal coverage across
the evaluation points. The simulations for projection coefficient showed the
same pattern: an approximately constant or approximately log-linear win ratio does not by itself justify a proportional win-fractions regression model, because the latter imposes the stronger
requirement of a time-invariant win ratio structure. Overall, the results
illustrate the tradeoff among the three approaches: the Mao--Wang and
PH copula estimators can gain substantial efficiency when their respective assumptions are correct, whereas the local IPCW estimator sacrifices some efficiency in exchange for robustness to these model assumptions.

\begin{table}[!htbp]
\centering
\caption{Simulation results for covariate-specific log win ratio estimation at grid points of the covariate.}
\label{tab:grid_simulation_main}
\scriptsize
\begin{tabular*}{\textwidth}{@{}rr@{\extracolsep{\fill}}rrrrrrrrrrrr@{}}
\toprule
& & \multicolumn{4}{c}{Mao--Wang estimator}
  & \multicolumn{4}{c}{Local IPCW estimator}
  & \multicolumn{4}{c}{PH copula estimator} \\
\cmidrule(lr){3-6} \cmidrule(lr){7-10} \cmidrule(lr){11-14}
Grid point & Truth
& Bias & SE & SEE & CP
& Bias & SE & SEE & CP
& Bias & SE & SEE & CP \\
\midrule
\multicolumn{14}{@{}l@{}}{\textit{proportional win fraction holds; PH copula model is correctly specified}}\\[0.25em]
-0.750
& 0.030
& -0.012 & 0.177 & 0.177 & 0.946
&  0.023 & 0.261 & 0.262 & 0.960
& -0.003 & 0.194 & 0.191 & 0.938 \\
-0.375
& 0.218
& -0.006 & 0.129 & 0.126 & 0.952
&  0.023 & 0.218 & 0.229 & 0.966
& -0.001 & 0.138 & 0.135 & 0.942 \\
0.000
& 0.405
& -0.000 & 0.118 & 0.111 & 0.938
& -0.007 & 0.237 & 0.229 & 0.942
&  0.002 & 0.121 & 0.118 & 0.946 \\
0.375
& 0.593
&  0.006 & 0.151 & 0.144 & 0.934
& -0.016 & 0.252 & 0.251 & 0.962
&  0.006 & 0.146 & 0.144 & 0.940 \\
0.750
& 0.780
&  0.012 & 0.209 & 0.203 & 0.938
&  0.012 & 0.291 & 0.316 & 0.964
&  0.010 & 0.190 & 0.191 & 0.938 \\
\midrule
\multicolumn{14}{@{}l@{}}{\textit{proportional win fraction  fails; PH copula model is correctly specified}}\\[0.25em]
-0.750
& 0.066
&  0.139 & 0.172 & 0.185 & 0.914
&  0.012 & 0.260 & 0.261 & 0.942
&  0.000 & 0.194 & 0.198 & 0.954 \\
-0.375
& 0.216
&  0.007 & 0.123 & 0.131 & 0.970
&  0.002 & 0.230 & 0.227 & 0.950
&  0.003 & 0.135 & 0.138 & 0.958 \\
0.000
& 0.330
& -0.087 & 0.101 & 0.105 & 0.890
& -0.005 & 0.221 & 0.228 & 0.964
&  0.005 & 0.119 & 0.121 & 0.962 \\
0.375
& 0.403
& -0.140 & 0.120 & 0.125 & 0.806
& -0.001 & 0.252 & 0.249 & 0.960
&  0.006 & 0.142 & 0.143 & 0.960 \\
0.750
& 0.436
& -0.153 & 0.168 & 0.176 & 0.862
&  0.006 & 0.324 & 0.311 & 0.952
&  0.008 & 0.182 & 0.184 & 0.952 \\
\midrule
\multicolumn{14}{@{}l@{}}{\textit{proportional win fraction  fails; PH copula model is misspecified}}\\[0.25em]
-0.750
& 0.508
& -0.151 & 0.164 & 0.178 & 0.876
& -0.008 & 0.286 & 0.296 & 0.964
& -0.081 & 0.183 & 0.199 & 0.934 \\
-0.375
& 0.331
&  0.013 & 0.119 & 0.127 & 0.974
&  0.008 & 0.239 & 0.239 & 0.950
&  0.093 & 0.138 & 0.148 & 0.932 \\
0.000
& 0.282
&  0.048 & 0.102 & 0.106 & 0.936
&  0.018 & 0.220 & 0.230 & 0.968
&  0.139 & 0.116 & 0.121 & 0.794 \\
0.375
& 0.382
& -0.065 & 0.125 & 0.128 & 0.930
&  0.004 & 0.237 & 0.244 & 0.960
&  0.037 & 0.133 & 0.135 & 0.940 \\
0.750
& 0.625
& -0.322 & 0.173 & 0.179 & 0.586
& -0.010 & 0.324 & 0.308 & 0.942
& -0.210 & 0.180 & 0.182 & 0.774 \\
\bottomrule
\end{tabular*}
\vspace{-0.1em}
\begin{minipage}{\textwidth}
\footnotesize
SE: empirical standard error; SEE: average estimated standard error; CP: empirical coverage probability of the 95\% confidence interval.
\end{minipage}
\end{table}

\begin{table}[!htbp]
\centering
\caption{Simulation results for estimation of linear coefficients in the covariate-specific log win ratio.}
\label{tab:simulation_main}
\scriptsize
\setlength{\tabcolsep}{2.7pt}
\begin{tabular*}{\textwidth}{@{}ll@{\extracolsep{\fill}}rrrrrrrrrrrr@{}}
\toprule
& &
\multicolumn{4}{c}{Mao--Wang estimator}
& \multicolumn{4}{c}{Local IPCW estimator}
& \multicolumn{4}{c}{PH copula estimator} \\
\cmidrule(lr){3-6}
\cmidrule(lr){7-10}
\cmidrule(lr){11-14}
Coefficient & Truth
& Bias & SE & SEE & CP
& Bias & SE & SEE & CP
& Bias & SE & SEE & CP \\
\midrule

\multicolumn{14}{@{}l@{}}{\textit{Constant $\mathrm{WR}(X)$; proportional win fraction holds}}\\[0.25em]

Intercept
& 0.405
& -0.002 & 0.100 & 0.107 & 0.964
& 0.007 & 0.155 & 0.157 & 0.956
& -0.001 & 0.102 & 0.109 & 0.968 \\

$X_1$ slope
& $0$
& 0.000 & 0.116 & 0.112 & 0.936
& 0.006 & 0.247 & 0.249 & 0.952
& 0.003 & 0.112 & 0.103 & 0.914 \\

$X_2$ slope
& $0$
& 0.000 & 0.120 & 0.113 & 0.932
& -0.014 & 0.246 & 0.248 & 0.960
& 0.000 & 0.105 & 0.103 & 0.948 \\

\midrule

\multicolumn{14}{@{}l@{}}{\textit{Approximately constant $\mathrm{WR}(X)$; proportional win fraction fails}}\\[0.25em]

Intercept
& 0.407
& -0.045 & 0.114 & 0.112 & 0.918
& 0.002 & 0.158 & 0.159 & 0.954
& 0.000 & 0.135 & 0.135 & 0.952 \\

$X_1$ slope
& -0.039
& 0.006 & 0.121 & 0.118 & 0.944
& 0.000 & 0.249 & 0.250 & 0.972
& 0.006 & 0.110 & 0.108 & 0.940 \\

$X_2$ slope
& 0.027
& -0.008 & 0.114 & 0.116 & 0.946
& 0.002 & 0.252 & 0.252 & 0.950
& -0.003 & 0.109 & 0.111 & 0.952 \\

\midrule

\multicolumn{14}{@{}l@{}}{\textit{Log-linear $\mathrm{WR}(X)$; proportional win fraction holds}}\\[0.25em]

Intercept
& 0.405
& 0.000 & 0.105 & 0.111 & 0.962
& 0.007 & 0.157 & 0.158 & 0.952
& 0.001 & 0.106 & 0.111 & 0.964 \\

$X_1$ slope
& 0.500
& 0.006 & 0.126 & 0.121 & 0.938
& -0.026 & 0.252 & 0.252 & 0.950
& 0.009 & 0.118 & 0.106 & 0.902 \\

$X_2$ slope
& 0.250
& 0.003 & 0.120 & 0.115 & 0.928
& -0.022 & 0.247 & 0.249 & 0.960
& 0.004 & 0.104 & 0.102 & 0.936 \\

\midrule

\multicolumn{14}{@{}l@{}}{\textit{Approximately log-linear $\mathrm{WR}(X)$; proportional win fraction fails}}\\[0.25em]

Intercept
& 0.496
& -0.122 & 0.113 & 0.112 & 0.778
& 0.001 & 0.159 & 0.162 & 0.958
& 0.005 & 0.136 & 0.137 & 0.958 \\

$X_1$ slope
& 0.381
& -0.089 & 0.115 & 0.113 & 0.874
& -0.014 & 0.249 & 0.256 & 0.962
& 0.015 & 0.113 & 0.114 & 0.940 \\

$X_2$ slope
& 0.261
& -0.018 & 0.115 & 0.116 & 0.946
& -0.001 & 0.263 & 0.257 & 0.942
& 0.003 & 0.112 & 0.114 & 0.944 \\

\bottomrule
\end{tabular*}

\vspace{-0.1em}
\begin{minipage}{\textwidth}
\footnotesize
SE: empirical standard error; SEE: average estimated standard error;
CP: empirical coverage probability of the 95\% confidence interval.
\end{minipage}
\end{table}

\begin{figure}[htbp]
\centering
\includegraphics[width=\textwidth]{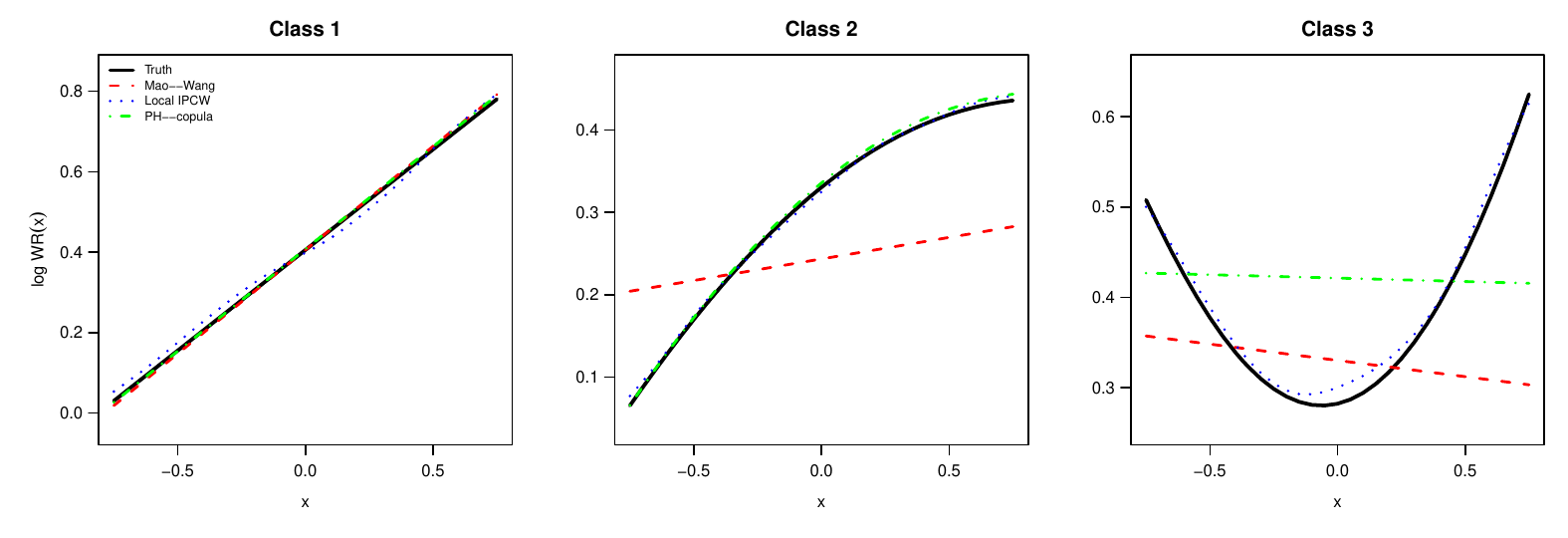}
\caption{Mean pointwise estimates of the covariate-specific log win ratio for the three classes of data generating processes.}
\label{fig-1}
\end{figure}

The results in Table~\ref{tab:simulation_test} show that rejection rates
under the near-null settings are generally close to the nominal
0.05 level, with some modest elevation in the two-dimensional setting.
Under nonlinear alternatives, power increased with sample size and with
the magnitude of departure from log-linearity. The  test based on PH copula model was
much more powerful than the local IPCW test when the working PH
and copula models were correctly specified, reflecting the efficiency
gained from the model-based structure. The local IPCW test was less
powerful, particularly with two covariates, but avoids the PH and copula
working-model restrictions.

\begin{table}[!htbp]
\centering
\caption{Rejection rates for tests of log-linearity of the win ratio function.}
\label{tab:simulation_test}
\small

\begin{minipage}{0.76\textwidth}
\centering

\begin{tabular}{llrrr@{\hspace{1.2em}}rrr}
\toprule
& &
\multicolumn{3}{c}{One covariate}
& \multicolumn{3}{c}{Two covariates} \\
\cmidrule(lr){3-5}\cmidrule(lr){6-8}
Method & Sample size
& $H_0$ & $H_{11}$ & $H_{12}$
& $H_0$ & $H_{11}$ & $H_{12}$ \\
\midrule
Local IPCW
& 300 & 0.040 & 0.148 & 0.366 & 0.026 & 0.098 & 0.304 \\
& 500 & 0.042 & 0.208 & 0.500 & 0.038 & 0.140 & 0.476 \\
& 800 & 0.052 & 0.286 & 0.694 & 0.066 & 0.194 & 0.664 \\
\midrule
PH--copula
& 300 & 0.064 & 0.968 & 1.000 & 0.066 & 0.834 & 1.000 \\
& 500 & 0.076 & 1.000 & 1.000 & 0.070 & 0.978 & 1.000 \\
& 800 & 0.056 & 1.000 & 1.000 & 0.080 & 1.000 & 1.000 \\
\bottomrule
\end{tabular}

\vspace{0.3em}

\footnotesize
\raggedright
\(H_0\) denotes the near-null log-linear setting. \(H_{11}\) and
\(H_{12}\) denote moderate and stronger departures from log-linearity,
respectively.

\end{minipage}
\end{table}

\section{Application in a lung cancer trial}
\label{sec:application}

We illustrate the proposed methods using CALGB 9633 (C9633), a randomized
phase III trial of adjuvant chemotherapy in patients with completely resected
stage IB non-small-cell lung cancer. In this trial, 344 patients with
pathologically confirmed T2N0 disease were randomized after lobectomy or
pneumonectomy to paclitaxel plus carboplatin or observation. The published
analysis, based on the composite outcome of disease-free survival, showed
favorable but not statistically significant treatment effects overall and
suggested possible heterogeneity by tumor size, with greater benefit among
patients with tumors at least 4 cm \citep{Strauss2008}. Because death and
recurrence or progression have a natural clinical hierarchy, we reanalyzed
the trial using a prioritized outcome that distinguishes terminal and
nonterminal events rather than collapsing them into a single composite
endpoint.

We defined time to death as the primary outcome and time to recurrence or
progression as the secondary outcome. Disease recurrence was assumed to occur
at the first visit at which recurrence was detected. The covariate-specific
win ratio was evaluated at the clinically meaningful time horizon
\(\tau=5\) years. Age and tumor size were considered as continuous covariates.
For each covariate, we evaluated the win ratio curve at 21 equally spaced
grid points between the empirical 5th and 95th percentiles. We estimated
these curves using both the local IPCW and PH copula methods. For the
local IPCW estimator, we also conducted secondary analyses stratified by
performance status. For each estimated log win ratio curve, we computed its best linear
projection. For the overall analyses, the adequacy of a linear representation
was assessed using both the local IPCW and PH copula log-linearity tests,
respectively, as described in Section~\ref{sec:test_linearity}. For the
performance status subgroup analyses, the log-linearity assessment was based
on the local IPCW procedure.

\begin{figure}[htbp]
\centering
\includegraphics[width=0.9\textwidth]{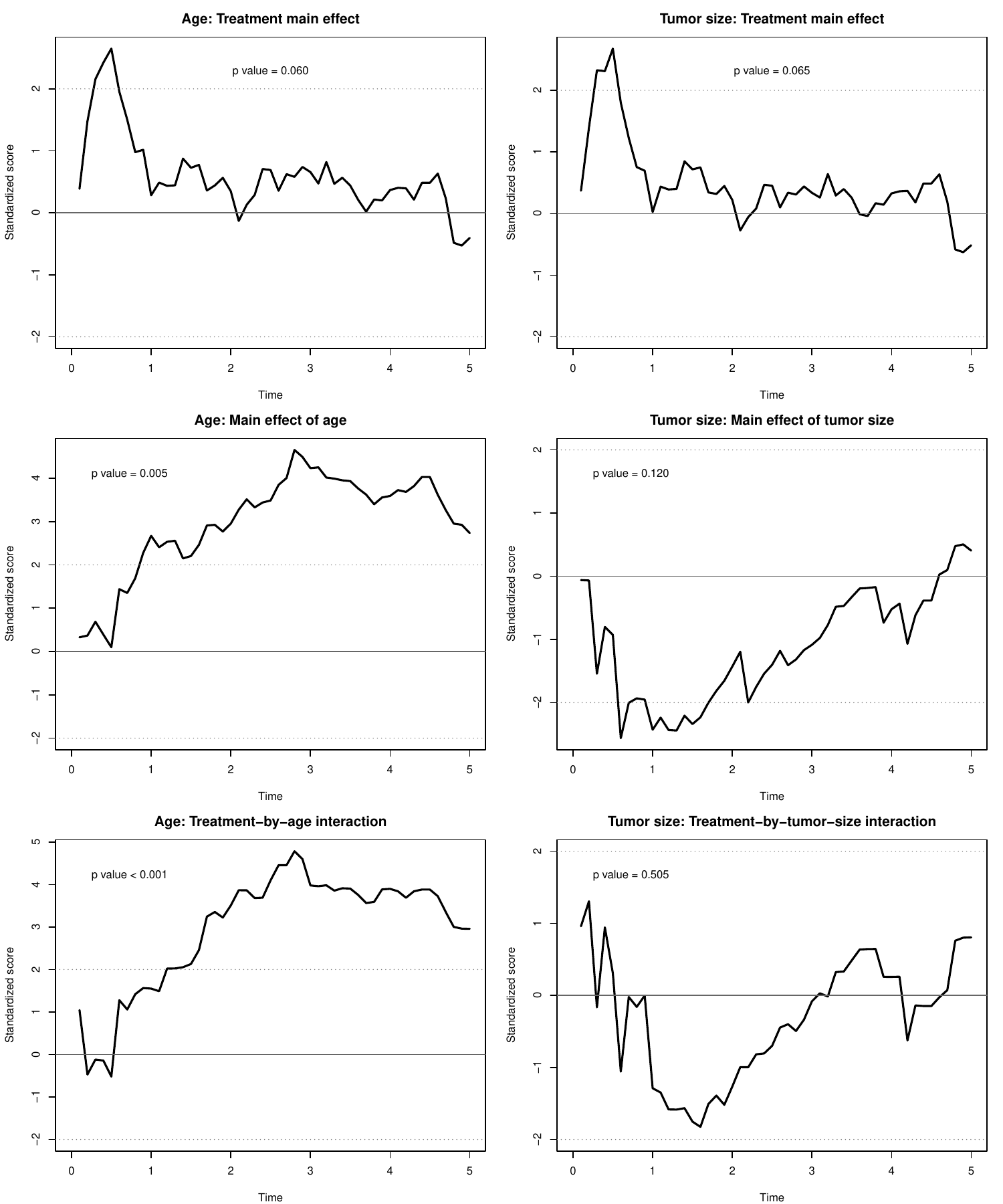}
\caption{
Score-process diagnostic plots for the proportional win-fractions assumption in the C9633 analysis. Within each analysis, the top, middle, and bottom panels assess the proportionality assumptions for the treatment main effect, the covariate main effect, and the treatment-by-covariate interaction, respectively. 
}
\label{fig:mw_diag}
\end{figure}

We first assessed the proportional win-fractions assumption using the
score-process diagnostics of \citet{MaoWang2021}. For age, the global
diagnostic rejected proportionality (\(p=0.005\)), with evidence of
nonproportionality for both the age main effect (\(p=0.005\)) and the
treatment-by-age interaction (\(p<0.001\)); see
Figure~\ref{fig:mw_diag}. For tumor size, the global diagnostic test did not
reject proportionality (\(p=0.175\)), nor was there evidence of
nonproportionality for the tumor-size main effect (\(p=0.120\)) or the
treatment-by-tumor-size interaction (\(p=0.505\)). These results suggest
that the proportional win-fractions assumption is questionable for age but
more compatible with tumor size. They also motivate examining the
covariate-specific win ratio at a fixed time horizon, which does not require
a time-invariant proportional win-fractions relationship.

Because the PH copula estimator additionally relies on proportional hazards
models for the marginal event time distributions, we also assessed the
marginal PH assumptions over the 5-year analysis horizon using Schoenfeld
residual diagnostics \citep{Schoenfeld1982,GrambschTherneau1994}. For age,
the corresponding \(p\) values for the death margin were 0.061 and 0.324 in
the observation and chemotherapy groups, respectively, and those for the
recurrence/progression margin were 0.145 and 0.386. For tumor size, the
corresponding \(p\) values were 0.193 and 0.496 for the death margin and
0.325 and 0.158 for the recurrence/progression margin. Thus, there was no
strong evidence against the marginal PH assumptions over the 5-year horizon
for either covariate. For the dependence between the two marginal event times, we used the two-parameter BB1 copula family \citep{Nelsen2006}.

\begin{figure}[htbp]
\centering
\includegraphics[width=0.9\textwidth]{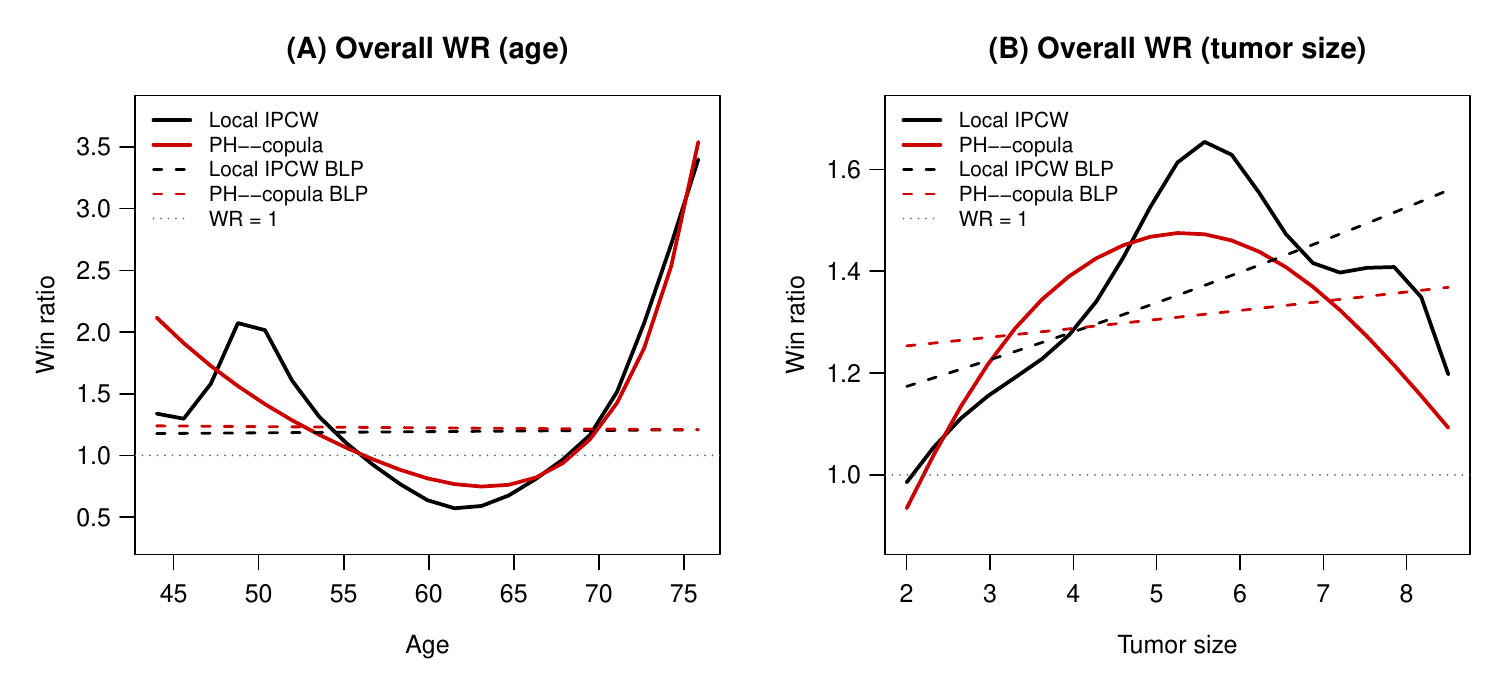}
\caption{
Estimated covariate-specific win ratio curves in C9633 at
\(\tau=5\) years. Panel (A) shows the win ratio as a function of age and
Panel (B) as a function of tumor size. The local IPCW and PH copula
estimates are shown together with the exponentiated best linear projections
of the corresponding log win ratio functions. Pointwise bootstrap standard
errors over the evaluation grids are reported in the Supplementary Material.
}
\label{fig:c9633_curves}
\end{figure}

The overall local IPCW and PH copula model-based results are summarized in
Table~\ref{tab:c9633_wr_age_tsize} and Figure~\ref{fig:c9633_curves}.
Pointwise estimates and bootstrap standard errors over the full evaluation
grids are reported in Tables~1 and 2 in the Supplementary Material.

\begin{table}[!htbp]
\centering
\footnotesize
\caption{Best linear projections and tests of log-linearity for
the covariate-specific win ratio in C9633.}
\label{tab:c9633_wr_age_tsize}

\setlength{\tabcolsep}{9pt}

\begin{tabular}{llccc}
\hline
Covariate / analysis & Method & Intercept (SE) & Slope (SE) & $p_l$ \\
\hline
\addlinespace[0.6em]
\multicolumn{5}{c}{\textbf{Age}} \\

Overall
& Local IPCW
& 0.127 (1.161)
& 0.00081 (0.01872)
& 0.0160
\\

Overall
& PH copula
& 0.250 (1.117)
& -0.00081 (0.01804)
& 0.0100
\\

PS = 0
& Local IPCW
& 0.966 (1.290)
& -0.00929 (0.02125)
& 0.0675
\\

PS = 1
& Local IPCW
& -0.050 (1.949)
& -0.00026 (0.03110)
& 0.0790
\\

\addlinespace[0.6em]

\multicolumn{5}{c}{\textbf{Tumor size}} \\

Overall
& Local IPCW
& 0.073 (0.419)
& 0.04369 (0.08644)
& 0.5815
\\

Overall
& PH copula
& 0.199 (0.430)
& 0.01352 (0.09056)
& 0.3540
\\

PS = 0
& Local IPCW
& 1.034 (0.588)
& -0.11328 (0.11196)
& 0.8495
\\

PS = 1
& Local IPCW
& -0.699 (0.612)
& 0.16396 (0.13075)
& 0.6755
\\

\hline
\end{tabular}

\vspace{0.5em}

\begin{flushleft}
\footnotesize
Intercept and slope refer to the best linear projection of
\(\log\{\mathrm{WR}(X)\}\), with bootstrap standard errors in parentheses.
The quantity \(p_l\) denotes the \(p\)-value for the log-linearity tests.
\end{flushleft}
\end{table}

For age, both approaches indicated that a simple log-linear form
was inadequate. The local IPCW best linear projection had intercept 0.127
and slope 0.00081, with bootstrap standard errors 1.161 and 0.01872,
respectively, and the log-linearity test rejected the log linear null
(\(p=0.016\)). The PH copula model-based analysis gave intercept 0.250 and slope
\(-0.00081\), with corresponding bootstrap standard errors 1.117 and
0.01804, and also rejected log-linearity (\(p=0.010\)). As shown in
Figure~\ref{fig:c9633_curves}, the two methods produced qualitatively
similar nonlinear age patterns, with substantially greater variation in the
estimated win ratio than is conveyed by their nearly flat best linear
projections. Thus, a near zero best linear slope should not be interpreted
as implying an approximately constant treatment effect over age.

The estimated age-specific curves should therefore be emphasized when
describing how treatment benefit varies with age. For the local IPCW
estimator, uncertainty was greater near the edges of the age distribution
because of reduced local information and boundary effects, so the behavior
at the youngest and oldest ages should be interpreted cautiously. The best
linear projection nevertheless remains useful as a low-dimensional summary
of the overall linear trend, even though the log-linearity test indicates
that it does not adequately capture the full age-specific pattern. 

For tumor size, the results were more compatible with a simple linear
summary, although the estimated curves showed some curvature. The local
IPCW best linear projection had intercept 0.073 and slope 0.04369, with
bootstrap standard errors 0.419 and 0.08644, respectively, and
log-linearity was not rejected (\(p=0.582\)). The PH copula model-based analysis gave
intercept 0.199 and slope 0.01352, with corresponding bootstrap standard
errors 0.430 and 0.09056, and likewise provided no evidence against
log-linearity (\(p=0.354\)). As shown in Figure~\ref{fig:c9633_curves},
both methods suggested an increase in the win ratio over the
lower-to-middle tumor-size range followed by some attenuation at larger
tumor sizes. Thus, unlike the age analysis, neither method provided strong
evidence that the best linear projection was inadequate.

The relative precision of the two estimators differed somewhat by
covariate. For age, the PH copula estimator had smaller pointwise bootstrap
standard errors throughout the evaluation grid. For tumor size, however,
the pointwise precision advantage was not uniform: the local IPCW estimator
was more precise at some lower tumor size values, whereas the PH copula
estimator was generally more precise over much of the middle and upper
range. The standard errors of the best linear projection coefficients were
also quite similar between the two methods. These results illustrate the
potential, rather than guaranteed, efficiency gain from the additional
model structure imposed by the PH copula estimator when its working models
are adequately specified. The full pointwise results are provided in
Tables 1 and 2 in the Supplementary Material. Secondary analyses stratified by performance status were conducted using the local IPCW estimator. For age, the best linear slopes were \(-0.00929\)
for \(PS=0\) and approximately zero for \(PS=1\), with log-linearity
\(p\)-values of 0.068 and 0.079, respectively. For tumor size, the
corresponding slopes were \(-0.11328\) and 0.16396, with \(p\)-values of
0.850 and 0.676. These subgroup analyses therefore provided little evidence
against a linear representation, although their interpretation should be
cautious because of the smaller subgroup sample sizes. The subgroup-specific
best linear projections are reported primarily as descriptive summaries of
the covariate trends.

Taken together with the proportionality diagnostics, these results illustrate
the different roles of the proposed procedures. For age, the proportional
win-fractions assumption was rejected, the local IPCW analysis rejected
log-linearity, and the model-based  test also rejected log-linearity. The covariate-specific curve therefore contains important information that is not captured by a single linear coefficient, although the best linear projection still provides a meaningful summary of the
overall linear trend. For tumor size, the proportional win-fractions diagnostic did not reject, and neither the local IPCW nor the PH copula analysis provided evidence
against log-linearity. The PH copula estimate also showed a relatively
smooth pattern over tumor size. In this setting, the best linear projection
provides a parsimonious and readily interpretable summary of the overall
relationship between tumor size and treatment benefit.

Overall, the C9633 analysis illustrates the complementary roles of the local
IPCW and PH copula procedures. At the 5-year horizon, the marginal PH
assumptions were reasonably compatible with the data, so the PH copula
estimator could be used alongside the more flexible local IPCW estimator and
could provide improved pointwise precision in some settings. Its
applicability, however, depends on the adequacy of the working marginal PH
models. When the analysis horizon was extended to \(\tau=10\) years, our analysis indicated that the
marginal PH assumption was rejected. The PH copula analysis was therefore
not supported at that horizon, whereas the local IPCW procedure, which does
not require the marginal PH assumption, remained applicable. We did not
pursue the 10-year analysis further. This illustrates that the local IPCW
approach is not merely a less structured alternative to a potentially more
efficient model-based estimator; in some settings it is the only one of the
two proposed procedures supported by the data.

\section{Discussion}

We developed a flexible framework for estimating the fixed-horizon
covariate-specific win ratio for prioritized right-censored time-to-event
outcomes. The local IPCW estimator avoids imposing a global regression
structure or time-invariant proportionality on the win-fraction scale and
therefore provides a flexible description of treatment-effect heterogeneity.
We also proposed a complementary PH copula model-based estimator that models the
component event time distributions and their dependence rather than imposing
a regression structure directly on the win ratio. When the working models
are adequately specified, this model-based approach can provide substantial
efficiency gains relative to the fully local estimator.

A central feature of the proposed framework is that flexible estimation can
be combined with interpretable low-dimensional summaries. The best linear
projection of the covariate-specific log win ratio summarizes its overall
linear trend without requiring the true function to be linear. The proposed
log-linearity tests provide formal assessments of whether this projection
offers an adequate representation of the full curve. The local IPCW test
avoids the PH copula working-model restrictions, whereas the model-based
test can have substantially greater power when those models are adequately
specified. When log-linearity is not rejected, the projection coefficients
can provide a parsimonious and readily interpretable summary. When
log-linearity is rejected, the projection remains meaningful as a best
linear summary, but the full covariate-specific curve is needed to describe
nonlinear features of treatment-effect heterogeneity.

The local IPCW estimator is best suited for a small number of prespecified,
clinically important effect modifiers, as is typical of nonparametric
smoothing methods. In higher-dimensional settings, it may be useful after
dimension reduction or through summaries on a prespecified low-dimensional
grid. The PH copula estimator provides an alternative when scientifically
reasonable models for the marginal event times and their
dependence are available. Its potential efficiency and power gains come at
the cost of greater sensitivity to misspecification of these working models.

The choice of copula family is an additional modeling consideration for the
PH copula model-based methods. In practice, one may restrict attention to a small set
of scientifically plausible copula families with different dependence
features, such as symmetric or asymmetric tail dependence, and compare their
fit using likelihood-based criteria such as AIC or BIC. Such comparisons
should not be interpreted as establishing the correctness of a particular
copula, since different copula families may fit the observed data similarly
while implying different forms of dependence. We therefore regard the copula
as a working model and recommend sensitivity analyses using several plausible
families when the estimated win ratio is materially affected by the dependence
specification. Formal copula goodness-of-fit testing and data-adaptive copula
selection are beyond the scope of this work.

Although we focused on two prioritized outcomes, the local framework is not
inherently limited to this setting. With more than two prioritized event
times, the win and loss probabilities can be decomposed into
priority-specific components and estimated using analogous IPCW
representations. We focused on two outcomes because this is the most common
setting in practice.

The present work assumes right-censored event times with recorded event
dates. In the C9633 analysis, recurrence was treated as occurring at the
first visit at which recurrence was recorded and was otherwise right censored
at the last disease assessment or death. When recurrence or progression is
known only to have occurred between scheduled assessments, a more rigorous
analysis would require extending the framework to interval-censored outcomes.
The survival function representation used here provides a natural starting
point for such an extension, but suitable interval-censored marginal and
joint estimators, together with their corresponding inference procedures,
would need to be developed. We therefore leave this extension for future research.

\section*{Supporting Information}

The Supplementary Material contains additional results for the real data analysis for the C9633 lung cancer trial and  proofs of the theoretical results.

\bigskip
\noindent
\textit{Address for correspondence:}\\
Zhiguo Li\\
Department of Biostatistics and Bioinformatics\\
Duke University\\
Durham, North Carolina, U.S.A.\\
Email: \texttt{zhiguo.li@duke.edu}


\begin{thebibliography}{99}

%\bibitem[Betensky et~al.(2002)]{BetenskyEtAl2002}
%Betensky, R. A., Lindsey, J. C., Ryan, L. M., and Wand, M. P. (2002).
%A local likelihood proportional hazards model for interval censored data.
%\emph{Statistics in Medicine}, 21, 263--275.

\bibitem[Beran(1981)]{Beran1981}
Beran, R. (1981).
Nonparametric regression with randomly censored survival data.
Technical Report, University of California, Berkeley.

\bibitem[Buyse(2010)]{Buyse2010}
Buyse, M. (2010).
Generalized pairwise comparisons of prioritized outcomes in the two-sample problem.
\emph{Statistics in Medicine}, 29, 3245--3257.


\bibitem[Chernozhukov(2025)]{Chernozhukov2025}
Chernozhukov, V., Demirer, M., Duflo, E., and Fern\'andez-Val, I. (2025).
Fisher--Schultz Lecture: Generic machine learning inference on heterogeneous
treatment effects in randomized experiments, with an application to immunization
in India.
\emph{Econometrica}, 93(4), 1121--1164.

\bibitem[Dabrowska(1987)]{Dabrowska1987}
Dabrowska, D. M. (1987).
Nonparametric regression with censored survival time data.
\emph{Scandinavian Journal of Statistics}, 14, 181--197.

\bibitem[Dong et~al.(2018)]{Dong2018}
Dong, G., Qiu, J., Wang, D., and Vandemeulebroecke, M. (2018).
The stratified win ratio.
\emph{Journal of Biopharmaceutical Statistics}, 28, 778--796.

\bibitem[Dong et~al.(2020)]{Dong2020}
Dong, G., Mao, L., Huang, B., Gamalo-Siebers, M., Wang, J., Yu, G., et al. (2020).
The inverse-probability-of-censoring weighting (IPCW) adjusted win ratio statistic: an unbiased estimator in the presence of independent censoring.
\emph{Journal of Biopharmaceutical Statistics}, 30, 882--899.

\bibitem[Dong et~al.(2023)]{Dong2023}
Dong, G., Hoaglin, D. C., Huang, B., Cui, Y., Wang, D., Cheng, Y., et al. (2023). 
The stratified win statistics (win ratio, win odds, and net benefit).
\emph{Pharmaceutical Statistics}, 22, 748--756.

\bibitem[Fan and Gijbels(1996)]{FanGijbels1996}
Fan, J. and Gijbels, I. (1996).
\emph{Local Polynomial Modelling and Its Applications}.
Chapman and Hall, London.


\bibitem[Grambsch and Therneau(1994)]{GrambschTherneau1994}
Grambsch, P. M. and Therneau, T. M. (1994).
Proportional hazards tests and diagnostics based on weighted residuals.
\textit{Biometrika}, 81, 515--526.


%\bibitem[Hernán and Robins(2020)]{HernanRobins2020}
%Hernán, M. A. and Robins, J. M. (2020). \emph{Causal Inference: What If}. Chapman and Hall/CRC, Boca Raton.


\bibitem[Lin and Ying(1993)]{LinYing1993}
Lin, D. Y. and Ying, Z. (1993).
A simple nonparametric estimator of the bivariate survival function under
univariate censoring.
\emph{Biometrika}, 80, 573--581.


\bibitem[Mielniczuk(1987)]{Mielniczuk1987}
Mielniczuk, J. (1987).
Some asymptotic properties of kernel estimators of a density function in case
of censored data.
\emph{Annals of Statistics}, 15, 766--773.

\bibitem[Mao and Wang(2021)]{MaoWang2021}
Mao, L. and Wang, T. (2021).
A class of proportional win-fractions regression models for composite outcomes.
\emph{Biometrics}, 77, 1265--1275.

\bibitem[Nelsen(2006)]{Nelsen2006}
Nelsen, R. B. (2006).
\emph{An Introduction to Copulas}.
2nd ed. Springer, New York.

\bibitem[Oakes(2016)]{Oakes2016}
Oakes, D. (2016).
On the win ratio statistic in clinical trials with multiple types of event.
\emph{Biometrika}, 103, 742--745.

\bibitem[Parner and Overgaard(2025)]{Parner2025}
Parner, E. T. and Overgaard, M. (2025).
Estimation of win, loss probabilities, and win ratio based on right-censored
event data.
\emph{Scandinavian Journal of Statistics}, 52, 170--184.

\bibitem[Péron et~al.(2018)]{Peron2018}
Péron, J., Buyse, M., Ozenne, B., Roche, L., and Roy, P. (2018).
An extension of generalized pairwise comparisons for prioritized outcomes in
the presence of censoring.
\emph{Statistical Methods in Medical Research}, 27, 1230--1239.

\bibitem[Pocock et~al.(2012)]{Pocock2012}
Pocock, S. J., Ariti, C. A., Collier, T. J., and Wang, D. (2012).
The win ratio: a new approach to the analysis of composite endpoints in
clinical trials based on clinical priorities.
\emph{European Heart Journal}, 33, 176--182.


\bibitem[Pr{\ae}stgaard and Wellner(1993)]{PraestgaardWellner1993}
Pr{\ae}stgaard, J. and Wellner, J. A. (1993).
Exchangeably weighted bootstraps of the general empirical process.
\textit{The Annals of Probability}, \textbf{21}, 2053--2086.

\bibitem[R Core Team(2026)]{RCoreTeam2026}
R Core Team (2026).
\emph{R: A Language and Environment for Statistical Computing}.
R Foundation for Statistical Computing, Vienna, Austria.

\bibitem[Schoenfeld(1982)]{Schoenfeld1982}
Schoenfeld, D. (1982).
Partial residuals for the proportional hazards regression model.
\textit{Biometrika}, 69, 239--241.

\bibitem[Silverman(1986)]{Silverman1986}
Silverman, B. W. (1986).
\emph{Density Estimation for Statistics and Data Analysis}.
 London: Chapman and Hall.

\bibitem[Strauss et~al.(2008)]{Strauss2008}
Strauss, G. M., Herndon, J. E., Maddaus, M. A., Johnstone, D. W., Johnson, E. A., Harpole, D. H., et al. (2008).
Adjuvant paclitaxel plus carboplatin compared with observation in stage IB non-small-cell lung cancer: CALGB 9633 with the Cancer and Leukemia Group B,
Radiation Therapy Oncology Group, and North Central Cancer Treatment Group Study Groups.
\emph{Journal of Clinical Oncology}, 26, 5043--5051.


\bibitem[Sun and Ding(2021)]{SunDing2021}
Sun, T. and Ding, Y. (2021).
Copula-based semiparametric regression method for bivariate data under
general interval censoring.
\textit{Biostatistics}, 22, 315--330.


%\bibitem[Tibshirani and Hastie(1987)]{TibshiraniHastie1987}
%Tibshirani, R. and Hastie, T. (1987).
%Local likelihood estimation.
%\emph{Journal of the American Statistical Association}, 82, 559--567.


%\bibitem[van der Vaart and Wellner(1996)]{VanderVaartWellner1996}
%van der Vaart, A. W. and Wellner, J. A. (1996). \emph{Weak Convergence and Empirical processes: With Applications to Statistics}. Springer, New York.

\bibitem[Wand and Jones(1995)]{WandJones1995}
Wand, M. P. and Jones, M. C. (1995).
\emph{Kernel Smoothing}.
Chapman and Hall, London.

\bibitem[Wang et~al.(2025)]{WangIPTW2025}
Wang, D., Zheng, S., Cui, Y., He, N., Chen, T., and Huang, B. (2025).
Adjusted win ratio using the inverse probability of treatment weighting.
\emph{Journal of Biopharmaceutical Statistics}, 35, 21--36.

\bibitem[Zhou et al.(2017)]{ZhouHuSun2017}
Zhou, Q., Hu, T. and Sun, J. (2017).
A sieve semiparametric maximum likelihood approach for regression analysis
of bivariate interval-censored failure time data.
\textit{Journal of the American Statistical Association}, 112, 664--672.


\end{thebibliography}
\end{document}